\documentclass[twocolumn]{aastex631}

\usepackage{physics,amsmath,amssymb,bbold,amsthm,mathrsfs}
\usepackage{xcolor,url}

\newcommand{\mki}{
  Kavli Institute for Astrophysics and Space Research, 
  Massachusetts Institute of Technology , 77 Massachusetts  Ave., 
  Cambridge, MA 02139, USA
}

\newcommand{\kms}{$\mathrm{km\,s}^{-1}$}

\newcommand{\Change}[1]{{\color{red}\bf #1}}
\renewcommand{\Change}[1]{#1}

\begin{document}

\title{Methods to Test the Source of the Extreme Gas Motions in WS 35}

\author[0000-0003-2602-6703]{Sean J. Gunderson} \email{seang97@mit.edu}\affiliation{\mki}

\author[0000-0002-7204-5502]{Richard Ignace}
\email{ignace@etsu.edu}\affiliation{Department
 of Physics \& Astronomy, East Tennessee State University, Johnson City, TN 37614 USA}

 \author[0000-0001-7946-1034]{Walter W. Golay}
\affiliation{Center for Astrophysics $|$ Harvard \& Smithsonian, 60 Garden St., Cambridge, MA 02138, USA}

\begin{abstract}

We present theoretical arguments toward the plausibility of \Change{a stellar wind} to explain the 16000\,\kms\ line broadening in the optical spectra of WS~35, the central star in the Pa~30 nebula. The wind model is discussed in the context of super-Eddington flows. \Change{We argue that WS\,35 potentially occupies a new regime of wind driving theory as the first metal-only wind. While this framework provides a promising avenue for explaining the high speed flow, questions remain about the source's true nature. We further describe how future radio observations can provide an independent test of the spherical wind scenario.  A magnetically channeled wind would likely produce a relatively flat and bright radio spectral energy distributions.  By contrast a spherical wind should result in a thermal radio spectrum with a canonical  continuum slope of $\nu^{0.6}$, and a brightness level consistent with the currently predicted mass-loss rate.}

\end{abstract}


\section{Introduction} \label{sec:intro}
Pa~30 is an exceptional supernova (SN) remnant \citep{Oskinova2020} displaying radial filaments in essentially a spherically symmetric distribution \citep{Fesen2023}. Pa~30 refers to the nebula formed by the SN explotion in 1181 AD as recorded in Chinese and Japanese records \citep{2002ISAA....5.....S}. In addition to the unique nebular morphology, at its center is the remnant object WS~35 (IRAS 00500+6713, WD J005311) that is possibly the result of a double-degenerate white dwarf (WD) merger \citep{Gvaramadze2019, Lykou2023}.  The resulting merger product has a surface temperature of 200\,kK and has been proposed to be either a Wolf-Rayet (WR) star, a WD, or a super-Chandrasehkar object \citep{Gvaramadze2019,Oskinova2020,Lykou2023}.

The optical and ultraviolet spectra arising from the central star show prominent emission lines with line broadening of up to $16000$\,\kms\ \citep{Gvaramadze2019,Lykou2023}. Interpreted in the form of a spherical wind, this extreme velocity is associated with the wind terminal speed $v_\infty$. The inferred wind mass flux $\dot{M} \approx 3\times 10^{-6}\,M_\odot$\,yr$^{-1}$ is high, but not extreme: it is comparable to that of the O-supergiant $\zeta$\,Puppis \citep{2017A&A...603A..86S,Cohen2020,Gunderson2024}. \Change{Note that with an observed luminosity of $\log(L_\mathrm{Obs}/L_\odot)\approx4.6$ (see Tab.~\ref{tab:StellarProps}), the luminosity in wind kinetic energy is larger than the radiant luminosity with $\dot{M}v_\infty^2/2 = 2 L_\mathrm{Obs}.$}

On the other hand, the line broadening of $16000$\,\kms\ is quite extreme for a wind.  It is more like the explosive speeds observed from supernovae \citep[e.g.,][]{2015MNRAS.451.1973S, 2024MNRAS.532.1887P}. The challenge in the case of WS~35 is explaining how a sustained wind can reach such high speeds.

The spectrum of WS~35 is replete with strong O and Ne lines of high ion states, similar to oxygen-rich WR (also known as WO) stars. WO stars have high speed winds up to around $5000$\,\kms\ \citep{1997A&A...325..178P}, which is at least $3\times$ slower than WS~35.  Line-driven winds of OB stars tend to produce terminal speeds of 2-3 times the surface escape speed of the star \citep[e.g.,][]{2024A&A...688A.105H}, but those wind terminal speed are generally \Change{slower than} $3000$\,\kms. 

The WD scenario has similar difficulties. Models of WD winds have concluded terminal velocities in the $v_\infty\approx1000-5000$\,\kms\ range \citep{MacDonald1992,Krticka2020}. This has been the main argument against the line broadening being caused by line-driven wind expansion. However, as we will discuss, radiation driving through continuum absorption can produce terminal velocity much larger than traditional line driving.

\Change{It has alternatively been suggested that the wind could be aided by an extensive magnetic field to achieve the requisite speeds or perhaps that there is a magnetosphere in corotation causing line broadening.} This was suggested by \citet{Gvaramadze2019} in response to the challenge of explaining such a high wind speed. In application to WS~35, imagine a dipole magnetic field with its axis aligned with the star's rotation axis \Change{and gas filling the magnetosphere out to the Alfv\'{e}n radius. If the Alfv\'{e}n radius is large, rapidly rotating, and the gas of sufficient optical depth in the line opacity, the line widths might be explained in terms of solid body the rotation of magnetosphere.}

For discussion purposes it is helpful to invoke the rigidly rotating magnetosphere (RRM) model of \cite{2005MNRAS.357..251T}. \Change{The fundamental quantity in the RRM model is the so called magnetic confinement parameter
\begin{equation}
    \eta_* = \frac{B_*^2 R_*^2}{\dot{M}v_\infty},\label{eq:etastar}
\end{equation}
where $B_*$ is the magnetic field measured at the magnetic equator \citep{2002ApJ...576..413U}. \cite{Lykou2023} placed an upper limit of 2.5\,MG on the surface field from the absence of Zeeman splitting in the \ion{O}{8} 6064\,\AA\, line. This field strength corresponds to a confinement parameter of $\eta_* \leq 2279$, assuming $v_\infty=16000$\,km\,s$^{-1}$ and the estimated $R_*=0.151\,R_\odot$ from \citet{Gvaramadze2019}, which is well within normal levels of magnetically strong OB stars \citep{Donati2006,Petit2019}.
}

\Change{While the magnetic field would be able to contain the gas, there are potential problems with mechanically aiding the wind through a magnetic field's rotation. If we assume solid body rotation out to the Alfv\'{e}n radius $r_\mathrm{A} = \eta_*^{1/4} R_* \approx 7 R_*$, the rotation rate of the system will be $0.02$\,s$^{-1}$. This would correspond to a stellar rotation velocity of $2300$\,km\,s$^{-1}$. The critical rotation velocity
\begin{equation}
    v_\mathrm{c} = \sqrt{\frac{G M_*}{R_*}},\label{eq:criticalspeed}
\end{equation}
on the other had, for WS\,35, assuming $M_* \sim M_\odot$, is only $v_\mathrm{c}\approx 1100$\,km\,s$^{-1}$. In other words, we would need the system to be rotating at twice the break up velocity of the central star for the wind to be aided by a rotating magnetosphere.

If we instead use the $100$\,MG field limit from \citet{Gvaramadze2019}, the stellar rotation velocity drops to $370$\,km\,s$^{-1}$. There are additional problems with this field limit, beyond the changes it would cause in line shapes; primarily the problem of magnetic braking. From \citet{udDoula2009}, the spin-down rate of a magnetic OB star is
\begin{equation}
    \tau = \frac{3}{2}\frac{k M_*}{B_* R_*}\sqrt{\frac{v_\infty}{\dot{M}}},
\end{equation}
where $k$ is the moment of inertia of the star (typically 0.1). For the $100$\,MG field, this would correspond to a $\tau \approx 27$\,yr spin-down time. This spin-down acts as an e-folding-time since WS\,35's supernovae in 1181\,CE
\begin{equation}
    v_\mathrm{rot} = v_0 \exp(-\frac{t - 1181}{\tau}),
\end{equation}
which then implies that the star's initially velocity was $v_\infty \sim 10^{16}$\,km\,s$^{-1}$. This is obviously an unphysical velocity, but it means that WS\,35 should have spun down already if its magnetic field is as strong as \citet{Gvaramadze2019} estimated or that its field is significantly weaker than $100$\,MG.}

\Change{The above discussion presents significant problems to the application of a magnetosphere to explain the line broadening. It is for this reason that we take the position that the source of the broadening is from a spherically expanding wind. The primary question then is how the wind can get to such extreme speeds. The concerns and arguments by \citet{Gvaramadze2019} are warranted; the scaling laws for WR winds do not apply to WS\,35. At the same time, there is also the question of how the presence of the wind could be tested since optical and UV data are inconclusive.}

\Change{The focus of this paper is to provide possible answers to both these questions.} In \S~\ref{sec:Theory}, we discuss the plausibility of the wind in the context of super-Eddington continuum driving. \Change{In particular, we focus on how the physical parameters of the system can already explain the measured velocity along with potential changes to previous modeling approaches.} In \S~\ref{sec:RadioSED}, we detail the analytical formulae of the thermal and non-thermal components of the radio SED of a spherical wind discuss how they can be used to determine the source of the line broadening. \Change{In \S~\ref{sec:RadioData}, we discuss the currently available radio data on WS\,35.} Finally, in \S~\ref{sec:conclusions}, we give our conclusions.

\section{Super-Eddington Wind Driving}\label{sec:Theory}

Optical, UV, and X-ray studies have measured the surface temperature of WS~35 to be $T_*\approx200$\,kK \citep{Gvaramadze2019,Oskinova2020,Lykou2023}. This is hot enough to ionize the abundant metals (C, O, and Ne) almost completely, removing the primary absorption lines responsible for line driving \citep{Poniatowski2022}. Higher $Z$ elements that would be less ionized at this temperature are also not abundant enough (see Table~\ref{tab:StellarProps}) to provide sufficient opacity for classical line driving like in OB and WR stars.

\begin{deluxetable*}{lccc}
    \tablecaption{Stellar parameters of WS 35 from multi-wavelength studies. \label{tab:StellarProps}}
    \tablehead{
        \colhead{Parameter} & \colhead{\citet{Gvaramadze2019}} & \colhead{\citet{Lykou2023}} & \colhead{\citet{Oskinova2020}}
    }
    \startdata
        $v_\infty$ (km\,s$^{-1})$ & $16000\pm1000$ & 15000 & --\\
        $\dot{M}$ ($10^{-6}\,M_\odot\,\mathrm{yr}^{-1}$) & $3.0\pm0.4$ & $\leq 4$ & $3.5\pm0.6$\\
        $M_*$ ($M_\odot$) & $<1.5$ & $1.2\pm0.2$, $1.45\pm0.2$ & -- \\
        $R_*$ ($R_\odot$) & $0.151^{+0.035}_{-0.013}$ & $\leq0.2$ & -- \\
        $T_*$ (kK) & $200_{-22}^{+12}$ & $200-280$ & $211_{-23}^{+40}$ \\
        $B_*$ (kG) & $<10^5$ & $<2500$ & -- \\
        $\log(L_\mathrm{Obs}/L_\odot)$ & $4.51\pm0.14$ & $4.48-4.78$ & $4.6\pm0.14$ \\
        $X_\mathrm{He}$ & $<0.1$ & $<0.4$ & -- \\
        $X_\mathrm{C}$ & $0.2\pm0.1$ & $<0.26$ & 0.15 \\
        $X_\mathrm{O}$ & $0.8\pm0.1$ & $\leq 0.7$ & 0.61 \\
        $X_\mathrm{Ne}$ & -- & $\leq0.5$ & $0.15\pm0.03$ \\
        $X_\mathrm{Mg}$ & -- & -- & $0.04\pm0.02$ \\
        $X_\mathrm{Si}$ & -- & -- & $0.06\pm0.04$ \\
        $X_\mathrm{S}$ & -- & -- & 0.04 \\
        $X_\mathrm{Fe}$ & 0.0016 & -- & --
    \enddata
\end{deluxetable*}

However, it is still possible for the opacity to be large enough for radiation to drive a wind. At 200\,kK, Fe's opacity increases significantly in what is known as the ``Fe opacity bump" \citep{Jeffery2006,Ro2019}. At the same time, the many ionization stages of Fe have thousands of potential absorption lines \citep{Poniatowski2022} that can overlap, creating a pseudo-continuum opacity. As a consequence, the opacity can be enhanced enough to push a star beyond the Eddington limit, which can be defined in terms of this opacity:
\begin{equation}
    \Gamma = \frac{\kappa L_*}{4\pi G M c} = \frac{\kappa}{\kappa_\mathrm{e}}\Gamma_\mathrm{e},
\end{equation}
where $\Gamma_e$ is the Eddington parameter for electron opacity $\kappa_e$ \citep{Owocki2015}. Specifically, we can see from this equation that even if the star's luminosity is below the Eddington luminosity $L_\mathrm{Edd}=4\pi GMc/\kappa_\mathrm{e}$ as defined for electron continuum absorption, the star can go above the Eddington limit, i.e., $\Gamma > 1$, for other sources of opacity such as line absorption.

The effects of the Fe opacity bump have been explored as the source of instabilities in the interiors of stars, but in these cases the super-Eddington flow is generally suppressed by the dynamics of the stellar interiors. \citep{Jeffery2006,Sundqvist2013,Owocki2015,Ro2019,Poniatowski2021,Poniatowski2022}

For WS 35, the Fe opacity bump is exposed, so a resulting super-Eddington flow could be the source of the extreme $16000$\,km\,s$^{-1}$ line broadening measured by \citet{Gvaramadze2019}. To test whether such a continuum driven wind can reach these speeds, we will use the general theory derived by \citet{Owocki2017}. This theory was originally developed for the case of additional energy deposition $\Delta\dot{E}$ into the atmosphere such that the total deposition rate exceeds the Eddington luminosity $L_* + \Delta\dot{E} > L_\mathrm{Edd}$. While our assumed case does not fit this, the \citet{Owocki2017} theory was intended to be general to any situation in which the system exceeds the Eddington limit. The results of this theory are independent on this additional $\Delta\dot{E}$ and only depend on the luminosity that escapes the system and is observed.

For example, the observed luminosity is a reduced fraction of $L_*$, the surface luminosity,
\begin{equation}
    \frac{L_\mathrm{Obs}}{L_*} = \frac{1}{1+\xi \Gamma}.\label{eq:Lobs}
\end{equation}
due to the amount of photons being absorbed to support a high mass-loss $\dot{M}$ against the force of gravity. This is also known as photon-tiring, parameterized as
\begin{equation}
    \xi = \frac{GM_*\dot{M}/R_*}{L_*}\label{eq:PhotTiring}.
\end{equation}
In our case, the reduction in the true luminosity is due to an assumed blanket of Fe-lines in the atmosphere. 

The terminal velocity of this super-Eddington wind scales with the escape velocity from the surface as
\begin{equation}
    \frac{v_\infty^2}{v_\mathrm{esc}^2} = \frac{\Gamma}{1+\xi\Gamma} - 1,\label{eq:TermVeloc}
\end{equation}
where $v_\mathrm{esc} = \sqrt{2GM_*/R_*}$. Note, though, that we can use Equation~\eqref{eq:Lobs} to simplify the terminal velocity scaling to
\begin{equation}
    \frac{v_\infty^2}{v_\mathrm{esc}^2} = \frac{\kappa L_\mathrm{Obs}}{4\pi G M_* c} - 1.\label{eq:vinfEq}
\end{equation}
This form of the \citet{Owocki2017} theory directly shows that while the mass-loss rate is needed to determine the true luminosity of the star, the terminal velocity is only determined by the observed luminosity.

The \Change{escape} velocity is of course dependent on the mass and radius of the star. This is an acute problem in applying Equation~\eqref{eq:vinfEq} to WS 35 because we do not have well constrained values for both. As shown in Table~\ref{tab:StellarProps}, when either the mass or radius is constrained, the other only has upper limits. If WS 35 is a WD, this problem would be solved using the mass-radius relation \citep{Ezuka1999}, but the WR signatures in the optical spectra \citep{Gvaramadze2019,Lykou2023} suggest it is not.

We can avoid this degeneracy in the \Change{escape} velocity by noting that WS 35 is in the limit of $v_\infty \gg v_\mathrm{esc}$. This simplifies Equation~\eqref{eq:vinfEq} to
\begin{equation}
    v_\infty^2 \approx \frac{\kappa}{2\pi c} \frac{L_\mathrm{Obs}}{R_*}.
\end{equation}
This relation shows directly that a super-Eddington flow's final kinetic energy \Change{per gram} is independent of the mass holding on to it and the physical mechanism(s) \Change{that enhance the opacity. The wind only cares that the opacity is enhanced to begin with.} For our interest, it gives us a way of computing the required physical parameters necessary for achieving the proposed terminal velocity.

Using the best-fit radius and luminosity from \citet{Gvaramadze2019}, the wind opacity would need to be $\kappa \approx 40$\,cm$^2$\,g$^{-1}$ to achieve the quoted 16000\,\kms\ terminal velocity, or about 200 times that of the electron opacity in a H-depleted plasma. The Rosseland mean opacity for the Fe opacity bump inside an OB star, with solar-like abundances, is ten times lower \citep{LePennec2014} than this value. \Change{This presents a challenge for explaining how the requisite opacity could be achieved without serious Fe abundance enhancement. We argue that is highlights the real problem instead: current models are \textit{insufficient} to explain this rather extreme system. To be consistent with the measurements for WS~35, we identify two specific assumptions that may need modifications: metal abundances and local Doppler broadening.}

Metal abundances and gradients have a long history in explaining the radiation driving in massive stars (see \citealp{Poniatowski2022} for a modern formulation and the citations therein). \Change{WS\,35 present a new chapter in this discussion due to being both H and potentially He deficient. These two elements act as the brakes on the acceleration in normal radiation driven winds \citep{Springmann1992,Gayley1994}. Without these heavier particles\footnote{Heavy compared to the electrons that are pulled along by Dreicer fields.}, the wind would be able to accelerate with much less inertial drag. At the same time, the velocity gradient from accelerating reduces the Sobolev optical depth, further increasing the line force in a positive feedback loop. This feedback loop eventually saturates and is balanced by mass-loading with the prior mentioned H and He, but without these, the feedback could be allowed to grow to such a point that 16000\,km\,s$^{-1}$ is achieved.}

\Change{As an illustration, let us consider a toy model star that has burned all its H and He and has the same metal abundances as WS\,35. We can predict this star's terminal velocity using the recently released Line-driven Iterative Mass-loss Estimator\footnote{\url{https://lime.ster.kuleuven.be/}} \citep[LIME;][]{Sundqvist2025}. This star will have the same observed luminosity as WS\,35 in Table~\ref{tab:StellarProps}, but due to current limitations in LIME, we can only explore temperatures up to $T_\mathrm{eff}=60$\,kK. To compensate, we give the star a mass of $M_*=10\,M_\odot$ so that we are effectively considering a WO/WC star whose observed luminosity is reduced as was considered above for the super-Eddington wind. The wind of this star would have a $v_\infty = 17632$\,km\,s$^{-1}$ terminal velocity. This shows the feasibility of line driving models replicating the observed line broadening, but will require line lists and abundance tables to be reformulated for extremely hot and metal only objects.}

\Change{For WS\,35 in particular, the question is whether the system is He-depleted, or at least sufficiently He-poor. We highlight the possible supporting evidence for this from the discovery of the filamentary SN remnant by \citet{Fesen2023}. The measured He abundances from \citet{Gvaramadze2019} and \citet{Lykou2023} are potentially from the foreground gas in the remnant component. If so, WS\,35 would then be a metal-only object and the first observational example of a metal-only wind.}

\Change{On the other hand, the assumed value of 250\,km\,s$^{-1}$ for the local Doppler broadening used by \citet{Gvaramadze2019} may be an underestimate. \citet{Gayley1995} showed that significant Doppler broadening of absorption lines can significantly enhance the momentum deposition in radiation-driven winds. At sufficiently high values, the broadening would cause such severe line overlap that the wind opacity becomes gray like electron scattering. Combined with the already present Fe opacity bump, this gray opacity could push the wind opacity to the elevated levels needed for a super-Eddington flow. However, such high turbulence may present its own problems and require further improvements to radiation driving theories.}

\Change{While our above discussion shows that a spherically expanding wind is possible, it does not answer the question of whether the observed properties are from a wind. The previous optical, UV, and X-ray studies, through their modeling, require some level of assumptions about the morphology of the gas to interpret. What is needed is a more direct spectral diagnostic that can distinguish a spherical wind. For that, we turn to the radio band.}

\section{Wind Radio SED}\label{sec:RadioSED}

For the case of radio emission from a spherically expanding wind, we will use the radio SED model from \citet{Erba2022}. This model assumes a spherically symmetric, magnetized, and ionized wind. Given the morphology of the surrounding nebula, the spherical symmetry is potentially not a good assumption but allows us to use analytic expressions for the radio SED. Future radio observations will be able to confirm/reject this assumption.

The radio SED from \citet{Erba2022} takes the form
\begin{eqnarray}
    \frac{S_\nu}{S_0} = &&\left(\frac{\nu}{\nu_0}\right)^{0.6}+K_0\left(\frac{\nu_0}{\nu}\right)^{0.5}\nonumber\\
    &&\times\int_{-1}^{1}\int_{0}^{1}u^{m-1/2}\mathrm{e}^{-\tau_\nu(u,\mu)}\mathrm{e}^{-\nu_R(u)/\nu}\dd u\dd\mu,\label{eq:WindSED}
\end{eqnarray}
where $\nu_0 = 30\,\mathrm{GHz}$, $S_0$ is the flux at $\nu_0$, and $u=R_*/r$ is the inverse radial coordinate. The first portion of the expression accounts for the thermal radio emission \citep[e.g.,][]{Leitherer1991} component. The second part accounts for the nonthermal synchrotron emission \citep{Dougherty2003,Rybicki1986} from relativistic electrons, whose density distribution is described by $m$, a power-law slope free parameter.

The integrals in Equation~\eqref{eq:WindSED} account for the Razin effect \citep{Dougherty2003} that suppresses the synchotron emission. The Razin effect is frequency dependent, being strongest below the Razin frequency
\begin{equation}
    \nu_R(u) = 200\,\mathrm{GHz} \left(\frac{\mathrm{kG}}{B_*}\right)\left(\frac{n_0}{10^{13}\,\mathrm{cm}^{-3}}\right)u,
\end{equation}
where $n_0 = \dot{M}/(4\pi\mu_\mathrm{tot} m_\mathrm{H}R_*^2v_\infty)$ is the wind's fiducial number density. Here $\mu_\mathrm{tot}$ is the overall mean molecular weight for both electrons and ions. The constant $K_0$ is 
\begin{eqnarray}
    K_0 = &&3.78\times10^6\left(\frac{C_*}{10^{10}\,\mathrm{cm}^{-3}}\right)\left(\frac{R_*}{R_\odot}\right)^{1/3}\left(\frac{B_*}{\mathrm{kG}}\right)^{3/2}\nonumber\\
    &&\times\left(\frac{Z_\mathrm{rms}^2}{\mu_\mathrm{e}\mu_\mathrm{i}}\right)^{-2/3}\left(\frac{n_0}{10^{13} \mathrm{cm}^{-3}}\right)^{-4/3},
\end{eqnarray}
where $C_*$ is a scale constant describing the number density of relativistic electrons at the the base of the wind. Here $Z_\mathrm{rms}$ is the root mean square ionic charge, and $\mu_\mathrm{e,i}$ are the mean molecular weights of the electrons and ions, respectively.

Assuming a constant terminal speed flow, relevant to the outer regions of the wind where radio emission occurs, the optical depth $\tau_\nu$ of free-free absorption between any point to the observer \citep{Erba2022} is
\begin{equation}
    \tau_\nu = \frac{\tau_0}{2}u^3 \left(\frac{\theta-\mu\sin\theta}{\sin^{3}\theta}\right),
\end{equation}
with fiducial optical depth
\begin{eqnarray}
    \tau_0 = &&1.39\times10^8 \left(\frac{Z_\mathrm{rms}^2}{\mu_\mathrm{e}\mu_\mathrm{i}}\right)g_\nu^{\mathrm{ff}} \left(\frac{\nu_0}{\nu}\right)^2\left(\frac{R_*}{R_\odot}\right)\nonumber\\
    &&\times\left(\frac{n_0}{10^{13}\,\mathrm{cm}^{-3}}\right)^2\left(\frac{10^4\,\mathrm{K}}{T}\right)^{3/2}.\label{eq:tau0}
\end{eqnarray}
The optical depth is dependent on the line of sight to the observer, given by the direction cosine $\mu = \cos\theta$. It is also dependent on the Gaunt factor $g_\nu^\mathrm{ff}$ \citep{Leitherer1991}. For our radio SED, we have included the wavelength dependence of the Gaunt factor $g_\nu \propto \nu^{-0.1}$, which produces the canonical 0.6 spectral index on the thermal component.\footnote{\citet{Erba2022} used $g_\nu \sim 1$ for illustrative purposes.}

Due to the wind becoming optically thick to radio absorption at different frequencies, the surface of $\tau_\nu=1$ will have a dependence on the frequency, which \citet{Erba2022} parameterized as
\begin{eqnarray}
    \frac{R_\nu}{R_*}= &&359 \left(\frac{Z_\mathrm{rms}^2}{\mu_\mathrm{e}\mu_\mathrm{i}}\right)^{1/3}\left(\frac{\nu_0}{\nu}\right)^{2/3}\left(\frac{n_0}{10^{13}\,\mathrm{cm}^{-3}}\right)^{2/3}\left(\frac{R_*}{R_\odot}\right)^{1/3}\nonumber\\
    &&\times\left(\frac{10^4\,\mathrm{K}}{T}\right)^{3/2}g_\nu^{1/3}
\end{eqnarray}
Here note that many of the quantities in this SED model will vary significantly from those provided in the examples in \citet{Erba2022} because WS~35 is H- and potentially He-depleted. The reduction of half the available electrons for radio emission has significant consequences for parameters that depend on $Z_\mathrm{rms}$ and $\mu_\mathrm{e,i}$, both of which we detail now.

Following the procedure from \citet{Schulz2019} for defining emission measures in H-depleted plasmas, the electron and ion densities can be normalized in terms of another element. This allows us to determine the mean molecular weights using the abundances given in Table~\ref{tab:StellarProps}. While \citet{Lykou2023} found a potentially significant He fraction in the system, this may be foreground gas from the SN remnant. The X-ray measurements from \citet{Oskinova2020} are a better probe of the abundances of the central star, so we will use these measurements. The resulting electron and ion mean molecular masses are $\mu_\mathrm{e} = 5.83$ and $\mu_\mathrm{i} = 16.52$, along with $Z_\mathrm{rms} = 2.46$ for the root mean square charge and a $\mu_\mathrm{tot} = 2.83$ total mean molecular mass.

With these pieces we can also give the flux normalization
\begin{equation}
    S_0(\nu_0=30\,{\rm GHz}) = 12.8\,\mu\mathrm{Jy} \left(\frac{n_0}{10^{13}\,\mathrm{cm}^{-3}}\right)^{4/3}\left(\frac{R_*}{R_\odot}\right)^{8/3},\label{eq:WindSEDNorm}
\end{equation}
which makes use of the $d=2.3$\,kpc distance to the system. We can see from this normalization the effect of the H-depletion as the loss of half the electrons significantly reduces the radio flux. Thus the wind flux, based on the frequency dependence in Equation~\eqref{eq:WindSED}, ranges from about 1\,$\mu$Jy at 0.1\,GHz to 100\,$\mu$Jy at 100 GHZ, depending on the specific parameters.

\begin{figure*}
    \centering
    \includegraphics[width=\linewidth]{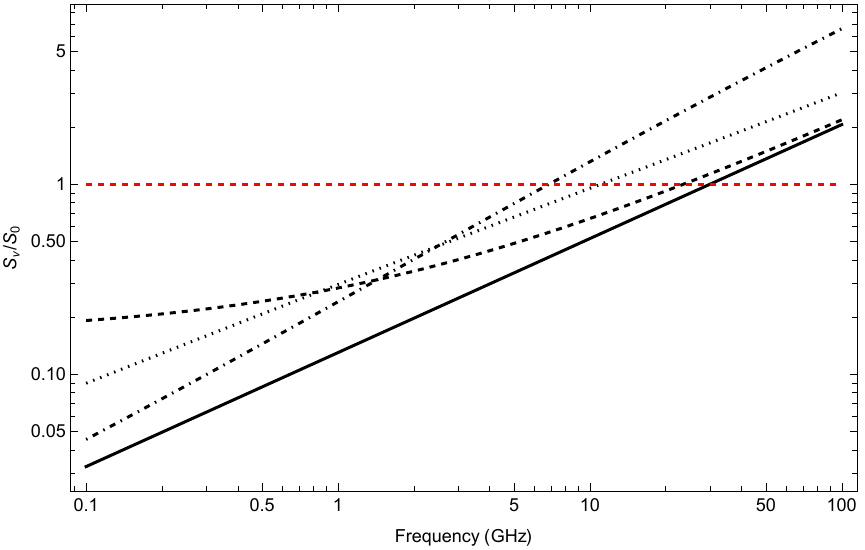}
    \caption{Possible slopes for the radio SED from a disk or wind emissions for different slope-determining parameters. The wind SEDs correspond to solid black (purely thermal; no synchrotron), dashed black ($m=0$), dotted black ($m=0.5$), dotted-dashed black is ($m=1$). The dashed red is the non-thermal magnetosphere emission of \cite{Leto2021}.}
    \label{fig:SEDcomparison}
\end{figure*}

In Figure~\ref{fig:SEDcomparison}, we plot the wind SEDs for different combinations of parameters that affect the slope. The black solid line is the case of a purely thermal wind with a slope of $0.6$. The other black lines show cases of differing non-thermal emission based on the distribution of relativistic electrons, parameterized by the slope $m$. Note that we specifically plot a normalized form of the SED $S_\nu/S_0$ because what we are focusing on is the non-zero slope.

\Change{In the same plot, the SED for the case of a magnetosphere from \citet{Leto2021}, who found a simple scaling relation for the luminosity from rotating magnetospheres
\begin{equation}
    L_\nu = 10^{13.6} \left(\frac{B_*}{\mathrm{kG}}\right)^{1.94}\left(\frac{R_*}{R_\odot}\right)^{3.88}\left(\frac{\mathrm{d}}{P_*}\right)^{1.94}\,\mathrm{ergs\,s}^{-1}\,\mathrm{Hz}^{-1}\label{eq:LetoLnu}
\end{equation}
fit a wide range of stellar types in the $\nu\approx0.1-400$\,GHz range. Here $P_*$ is the period of the star's rotation. The important feature of Equation~\eqref{eq:LetoLnu} is that it is frequency independent, so the SED of a magnetosphere would be a constant. This is shown in the red dashed line in Figure~\ref{fig:SEDcomparison}.}

\Change{While we provided highly critical arguments against the magnetosphere hypothesis in \S~\ref{sec:intro}, we concede that these were made from the upper limits of the magnetic field estimates. The limits from \citet{Lykou2023} provide particularly strong evidence against the magnetic field, but the number of uncertain quantities in the system leaves open many avenues for improving estimates.}

\Change{As such, we propose that radio observations can determine whether the system has a wind or a magnetosphere. Figure~\ref{fig:SEDcomparison} shows that the slope of the SED is uniquely determined by source of the emission.} Regardless of the strength of the Razin effect, a non-zero slope signal will immediately determine that the radio source is from an expanding wind. This in turn will answer the question of the source of the optical line broadening as being from the wind terminal velocity. Similarly, a constant (non-zero) flux measurement will be attributable to a magnetosphere.

\section{Very Large Array Observations}\label{sec:RadioData}
\Change{Existing archival observations place a deep limit on the absolute level of continuum radio emission. The Karl G. Jansky Very Large Array (VLA) observed WS\,35 under program VLA/22A-256 (PI: P. Zhou). The VLA collected a single two-hour continuum pointing in C-band (4--8\, GHz) on August 16, 2022, when the array was in D configuration. We reduced these observations using the VLA Pipeline v2024.1.1.22  in CASA v6.6.1 \citep{CASA:2022} and imaged the data using standard CASA {\tt tclean} parameters for wideband maximum-sensitivity imaging. We placed a strict 3$\sigma$ upper limit of 12\,$\mu$Jy at the position of WS\,35. From this non-detection, we can derive an upper limit to the mass loss rate of $\dot M<1.4\times10^{-4}\, M_\odot\,{\rm yr^{-1}}$ assuming a purely thermal radio component, an unsurprising result given the measured values are two orders of magnitude lower (see Table~\ref{tab:StellarProps}).

The existing non-detection provides an initial anchor point to measure the broadband spectral index. Although observations of WS\,35 exist at other radio frequencies, they suffer from sensitivities $\gtrsim10$x those of the archival C-band observation above and happen to all be lower frequency, where comparable or deeper sensitivities are required for useful combination with the C-band limit (see Figure~\ref{fig:SEDcomparison}). We reduced an L-band (1--2\,GHz) observation from the same program VLA/22A-256 using the same procedure described above, and found a far-weaker continuum 3$\sigma$ upper limit of 230\,$\mu$Jy. The VLA Sky Survey \citep[VLASS,][]{Lacy:2020} observed the field of WS\,35 on three occasions (most recently on March 16, 2023), and a source was not detected in a three-epoch stacked image, consistent with a 3$\sigma$ upper limit of $\lesssim 310$\,$\mu$Jy in S-band (2--4\,GHz). These more recent observations were all collected with the highly sensitive Extended VLA \citep[EVLA,][]{Perley:2011}. The position of WS\,35 was also observed by the NRAO VLA Sky Survey \citep[NVSS,][]{Condon:1998} on November 23, 1993, and was not detected to a 3$\sigma$ upper limit of $\lesssim 1.5$\,mJy in the narrower pre-EVLA upgrade L-band (1.4\,GHz). Given the scalings in Figure~\ref{fig:SEDcomparison}, high-frequency radio observations above C-band (4--8\,GHz) and millimeter observations of even somewhat comparable depth could provide a much more robust measurement of the radio SED.}

\section{Conclusions} \label{sec:conclusions}

Theoretical arguments have been presented for the \Change{applicability of radiation driving to explain the extreme velocity broadening measured in the optical spectra of WS\,35}. Prior work has brought attention to the difficulty in a radiation-driven wind reaching the proposed $16000$\,\kms\ velocities. However, such a fast speed is plausibly explained in the limit of a super-Eddington wind from significant opacity enhancements, \Change{but this raises the challenge of needing to explain the origin of the opacity enhancements. One possibility is that local Doppler broadening could be underestimated by \citet{Gvaramadze2019}. Higher microturbulence could result in sufficient overlap of lines to increase the opacity sufficiently.}

\Change{At the same time, it is worth considering WS\,35 to be in regimes as-yet-unexplored by radiation driving. This star may present the first example of a metal-only wind. The lack of inertial drag by H and He atoms owing to their absence permits higher wind velocities.  Current line-driving models can reproduce the observed wind high speeds of WS\,35, showing efficacy in the wind-driving hypothesis. However, achieving a wind speed of $\sim 16,000$ km/s involved only a toy model that was unable to reproduce other properties of WS\,35. While tantalizing, a more detailed theoretical investigation is needed.}

\Change{Our conclusion that the line broadening is caused by a wind is based also on the rejection of magnetic mechanisms. From the estimated field strengths and stellar parameters from Table~\ref{tab:StellarProps}, a rotating magnetosphere would need to violate the critical rotation rate of the central star either currently or at some point since 1181\,CE.}

\Change{However, we cannot fully rule out the possibility of a magnetosphere owing to a number of uncertain parameters. 
Radio measurements are therefore proposed as a test for the nature of the circumstellar environment.} For the wind case, the model developed by \citet{Erba2022} was used, which takes into account both thermal emission and a simplistic treatment for synchrotron emission, including the Razin effect.  A purely thermal radio emission would be consistent with a $\nu^{0.6}$ continuum slope. However, for a magnetosphere case, using the empirical non-thermal flux scaling from \citet{Leto2021} was used, a relatively flat SED over the entire range of considered frequencies would result. 

The shape of the radio SEDs, given in Figure~\ref{fig:SEDcomparison}, shows that this can be determined since even the most shallow wind SED is distinguishable from being constant. Radio observations across a wide range of frequencies should be undertaken. The primary limitation of this effort will be measuring the wind component due to the flux level predicted by Equation~\eqref{eq:WindSEDNorm}. The \textit{VLA} is capable of measuring this flux at 30\,GHz, but the atmospheric constraints at these high frequencies will pose difficulties. Longer observationally at low frequencies may be preferable to provide atmospheric condition to reach high statistical measurements. Doing so will reveal the nature of WS~35 and interesting implication studies of stellar evolution and mergers.

\begin{acknowledgments}

Support for SJG was provided by NASA through the Smithsonian Astrophysical Observatory (SAO) contract SV3-73016 to MIT for Support of the Chandra X-Ray Center (CXC) and Science Instruments. CXC is operated by SAO for and on behalf of NASA under contract NAS8-03060. 

The National Radio Astronomy Observatory is a facility of the National Science Foundation operated under cooperative agreement by Associated Universities, Inc.

We thank a number of individuals for their varied discussions on seemingly disparate topics for this paper, in alphabetical order: D. Debnath, D. Dong, R. Fesen, N. Moens, A. Sander, and J. Sundqvist.

Finally, we thank our referee K. G. Gayley for their thorough and detailed comments on our analysis. Their suggestions and questions significantly changed the conclusions of our theory and rose to the level of co-authorship.

\end{acknowledgments}

\bibliography{bib}{}
\bibliographystyle{aasjournal}

\end{document}